\documentclass[twoside]{article}

\evensidemargin\oddsidemargin
\advance \baselineskip 0pt plus 0.1pt minus 0.1pt
\input epsf
\usepackage[english]{babel}

\usepackage{latexsym}
\usepackage{amsbsy}
\usepackage{amsfonts}
\usepackage{amssymb}

\begin{document}
\def\Z{\hbox{{\sf Z}\kern-0.4em {\sf Z}}}

\title{Energy Bounds of Linked Vortex States}
\author{A.P. Protogenov\footnote{e-mail: alprot@appl.sci-nnov.ru}, V.A. Verbus \\
\\
{\fontsize{10pt}{12pt}\selectfont
{\em 
Institute of Applied Physics of the RAS, 603950 Nizhny Novgorod
}\/}\\
{\fontsize{10pt}{12pt}\selectfont
{\em 
Institute for Physics of Microstructures of the RAS, 603950 Nizhny Novgorod
}}\/
}

\maketitle

\begin{abstract}
Energy bounds of knotted and linked vortex states in a charged
two-component system are considered. It is shown that a set of
local minima of free energy contains new classes of universality.
When the mutual linking number of vector order parameter vortex lines 
is less than the Hopf invariant, these states have lower-lying energies.    
\end{abstract}

A tangle of vortex filaments is a system which attract attansion due to several reasons.
Along with the coherent state, which is the background of this vortex field
distributions, the filament system also contains a disorder combination due
to free motion of its fragments and a topological order because of the effects of
knotting and linking \cite{Vol,Mm} of its separate parts.

The study of soft condensed matter, whose universal behavior is determined by 
topological characteristics, is recognized as one of the most challenging problems 
of modern condensed matter physics \cite{Kbmsds}. 
The aim of this paper is to find the energy bounds for vortex states
with a set of numbers determining the knotting and linkig degree of the fields 
that take part in the description of the coherent state. 
We will use the Ginzburg-Landau model
\begin{equation} 
F =\int d^{3}x \, \biggl[\sum_{\alpha} \frac{1}{2m} \left| \left(\hbar \partial_{k} + 
i \frac{2e}{c} A_{k}\right) \Psi_{\alpha} \right|^2  +   
\sum_{\alpha}
\left(-b_{\alpha}|\Psi_{\alpha }|^{2} + 
\frac{c_{\alpha }}{2}|\Psi_{\alpha }|^4 \right) + \frac{\bf B^{2}}{8\pi} \biggr]
\end{equation}
with a two-component order parameter
\begin{equation}
\Psi_\alpha = \sqrt{2m} \,\rho \,\chi_\alpha , \, \, \, \,   
\chi_\alpha=|\chi_\alpha|e^{i \varphi_\alpha} \, , 
\end{equation}                                                                            
satisfying the $CP^1$ condition, $|\chi_{1}|^{2} + |\chi_{2}|^{2}=1$. 
This model is used in the context of the two-gap superconductivity \cite{Bfn} and in the non-Abelian 
field theory \cite{Fn1,Cho}. 

It has been shown in paper \cite{Bfn} that there exists an exact mapping of the model (1),
(2) into the following version of ${\bf n}$-field model:
\begin{equation}
F =\int d^{3}x \left[\frac{1}{4}\rho^{2}\left(\partial_{k}{\bf n}\right)^{2} + 
\left(\partial_{k}\rho \right)^{2} + \frac{1}{16}\rho^{2}{\bf c}^{2} + 
\left(F_{ik} - H_{ik}\right)^{2} + V(\rho, n_{3})\right] \, .
\end{equation}
To write down Eq.(3) dimensionless units and gauge invariant order parameter fields 
of the unit vector ${\bf n}={\bar \chi}{\boldsymbol{\sigma}}\chi$, 
where $\bar \chi = (\chi_{1}^{\ast}, \chi_{2}^{\ast})\,$, $\boldsymbol{\sigma}$ - Pauli matrices, 
and of the velocity  ${\bf c}={\bf J}/\rho^{2}$ have been used. 
The full current ${\bf J} = 2\rho^{2}({\bf j} - 4{\bf A})$ has a paramagnetic 
$\left({\bf j}=i[\chi_{1}\nabla \chi_{1}^{\ast} - c.c. + (1 \to 2)]\right)$ and 
a diamagnetic $(-4{\bf A})$ parts. Besides in Eq.(3) 
$F_{ik} = \partial_{i}c_{k} - \partial_{k}c_{i}$, $H_{ik} = {\bf n}\cdot[\partial_{i}{\bf n}\times \partial_{i}{\bf n}]:= \partial_{i}a_{k} - \partial_{k}a_{i} $. 

Setting in Eq.(3) ${\bf c} = 0$ we get Faddeev-Niemi model \cite{Fn2}. The numerical study of
the knotted configurations of ${\bf n}$-field in this model has been done 
in \cite{Gh,Bs,Hs}. The lower energy bound in this case
\begin{equation}
F \geqslant 32\pi^{2}\,|Q|^{3/4}
\end{equation}
is determined (\cite{Vk,Kr,Ward}) by the Hopf invariant, 
\begin{equation}
Q = \frac{1}{16\pi^{2}}\int d^{3}x \, \varepsilon_{ikl}a_{i}\partial_{k}a_{l} \, .
\end{equation}

At compactification $\Bbb R^{3} \to S^{3}$ and ${\bf n} \in S^{2}$, 
the integer $Q \in \pi_{3}(S^{2}) = \Bbb Z$ shows the degree of linking or knotting
of filamental manifolds, where the vector field ${\bf n}(x,y,z)$ is defined.  
In particular, for two linked rings (Hopf linking) Q=1, for the trefoil knot Q=6 and etc. 
Significant point is the following: 
$\pi_{3}(CP^{M}) = 0$ at $M > 1$ and $\pi_{3}(CP^{1}) = \pi_{3}(S^{2}) = \Bbb Z$ \cite{Aw}. 
In the latter case the order parameter (2) is two-component one \cite{Bfn} and linked or knotted 
soliton configurations are labeled by the Hopf invariant (5). 
In the $(3+0)D$ case of the free energy (3), Hopf invariant (5) is analogous to the Chern-
Simons action $(k/4\pi)\int dt\,d^{2}x\,\varepsilon_{\mu \nu \lambda}
a_{\mu}\partial_{\nu}a_{\lambda}$ determining 
strong correlations of $(2+1)D$ modes \cite{Apv,Pro}  
at semion value $k \simeq 2$. 

Let us assume that $\rho$ can be find from the minimal value of the potential $V(\rho)$, 
but the velocity ${\bf c}$ does not equal zero. Equation(3)
in this case has the following form: 
\begin{equation}
F = F_{n} + F_{c} - F_{int} = \int d^{3}x \left[\left(\left(\partial_{k}{\bf n}\right)^{2} + H_{ik}^{2}\right) + \left(\frac{1}{4}{\bf c}^{2} + F_{ik}^{2}\right) - 
2F_{ik}H_{ik}\right] \, .
\end{equation}

It is seen from Eq.(6) that a superconducting state with ${\bf c} \neq 0$ has the energy
which is less than the minimum in Eq.(4) 
due to renormalization of the coefficient $( = 1)$ in the second term of the functional $F_{n}$. 
To find the lower free energy bound in the superconducting state with ${\bf c} \neq 0$ 
we will use the following inequality: 
\begin{equation}
F_{n}^{5/6}\,F_{c}^{1/2} \geqslant (32\pi^{2})^{4/3}\,|L| \, \, , 
\end{equation}
where
\begin{equation}
L = \frac{1}{16\pi^{2}}\int d^{3}x \, \varepsilon_{ikl}c_{i}\partial_{k}a_{l} 
\end{equation} 
is the degree of mutual linking of the velocity ${\bf c}$ lines and of the 
magnetic field ${\bf H} = [\nabla \times {\bf a}]$ lines. It is also an integral of 
motion \cite{Zk}. 

The proof of the inequality (7) is based on employ of the H\"older inequality chain  
$|L| \leqslant ||{\bf c}||_{6}\cdot ||{\bf H}||_{6/5}$, 
$||{\bf H}||_{6/5} \leqslant ||{\bf H}||_{1}^{2/3} \cdot ||{\bf H}||_{2}^{1/2}$ 
and on the Ladyzhenskaya \cite{Lad}  
inequality \cite{Lad} $||{\bf c}||_{6} 
\leqslant 6^{1/6}||[\nabla \times {\bf c}]||_{2}$, , as well as on the results
of the paper \cite{Ward}. Here $||{\bf H}||_{p}:= 
(\int d^{3}x \, |{\bf H}|^{p})^{1/p}$. 
It follows also from the Schwartz-Cauchy-Bunyakovsky inequality that
\begin{equation}
2F_{n}^{1/2}\,F_{c}^{1/2} \geqslant F_{int} \, \, .  
\end{equation}
Setting the boundary value $F_{int}$ from Eq.(9) into Eq.(6), we get $F_{min}=(F_{n}^{1/2} - F_{c}^{1/2})^{2}$. Using the mimimum
value $32\pi^{2}|Q|^{3/4}$ of the function $F_{n}$ and the Eq.(7) we have finally
\begin{equation}
F  \geqslant 32\pi^{2}\,|Q|^{3/4}\,(1 - |L|/|Q|)^{2} \, \, .  
\end{equation}
The inequality (10) is the main result of the paper. The trivial case $Q=0$ 
should be considered after the limit $L=0$. Let us pay also an attention to the 
self-dual relation $F_{n} = F_{c}$ which follows from $F_{min}$.   

It follows from Eq.(10) that for all numbers $L < Q$ the energy of the ground state
is less than that in the model described in \cite{Fn2}, for which the inequality (4) is valid.
The origin of the energy decrease can be easily understood. Even under the conditions
of the existence of the paramagnetic part ${\bf j}$ of the current ${\bf J}$, 
the diamagnetic interaction in the superconducting state consumes its own current energy 
and a part of the energy relating to the ${\bf n}$-field dynamics for all state 
classes with $L < Q$. 

The case $\rho \neq const$ both for ${\bf c}=0$ \cite{Lnnt} and ${\bf c} \neq 0$ 
arouses certain interest. It is more complicated due to some reasons and will be considered 
in a separate paper. We only mention that in a soft case $\rho \neq const$ 
we need to have the proof of the compatibility of the peculiar value of the coefficient 
in r.h.s of Eq.(7) with stability condition \cite{Ward} with respect to dilatation transformations.
The equality in Eq.(12) under this remark should be understood as an 
ideal limit depending on topologial characteristics of knots and links only.  

In conclusion, we have found the energy bounds of the superconducting states 
using $CP^{1}$ version of Ginzburg-Landau model under the conditions of the
existence of linking and knotting phenomena of the ${\bf n}$- and ${\bf c}$-fields 
being the gauge invariant order parameters of the considered system. 
We have shown that the energy space of the local minima contains new state classes 
with $L < Q$. 

We would like to thank A.G. Abanov, L.D. Faddeev, E.A. Kuznetsov and G.E. Volovik
for advices, V.F. Gantmakher for the crucial remark, and G.M. Fraiman, A.G.
Litvak, V.A. Mironov for useful discussions. This work was supported in part
by the RFBR under the grant \# 01-02-17225.

\end{document}